\documentclass[prd,reprint,superscriptaddress]{revtex4-1}

\usepackage{amsmath,amssymb}
\usepackage{verbatim}
\usepackage{graphicx}
\usepackage{hyperref}
\usepackage{color}
\usepackage{mathrsfs}
\usepackage{slashed}

\usepackage[utf8]{inputenc}

\newcommand{\be}{\begin{equation}}
	\newcommand{\ee}{\end{equation}}
\newcommand{\bea}{\begin{eqnarray}}
	\newcommand{\eea}{\end{eqnarray}}

\newcommand{\ba}{\begin{array}}
	\newcommand{\ea}{\end{array}}

\def\double #1{#1{\hbox{\kern-2pt $#1$}}}

\newcommand{\bsubeq}{\begin{subequations}}
	\newcommand{\esubeq}{\end{subequations}}

\begin{document}
	
	\title{  Aspects of Non-Relativistic Supersymmetric Theories   }

		\author{Osman Ergec}
	\email{ergec24@itu.edu.tr}
	\affiliation{Department of Physics,
		Istanbul Technical University,
		Maslak 34469 Istanbul,
		T\"urkiye}

	\date{\today}

\begin{abstract}
Over the last decade, non-relativistic theories have attracted considerable attention. In general, such theories can be obtained by contracting relativistic parent theories. In this work, we discuss features of non-relativistic supersymmetric field theories from both the Galilean and Carrollian points of view that may be useful for constructing electric and magnetic non-relativistic theories.
\end{abstract}

\maketitle 
	 \textit{Introduction} --
Non-relativistic theories have received considerable attention for various motivations in recent years, both in the Carrollian \cite{Petkou:2022bmz, Duval_2014b, deboer2021carrollsymmetrydarkenergy, Bergshoeff:2023vfd, Grumiller:2025rtm, Kasikci:2023tvs, Kasikci:2023zdn, deBoer:2023fnj, Ecker:2024czx, Ruzziconi:2026bix, Bagchi:2025vri, Grumiller:2024dql, Baiguera:2022lsw, Duval_1994}  and Galilean \cite{Andringa:2010it, Bagchi:2014iea, Bergshoeff:2015uaa, Bergshoeff:2015ija, Son:2013rqa, Hartong:2022lsy, Baiguera:2023fus, Baiguera:2022cbp, Baiguera:2020jgy} settings. In general, one can consider two types of contractions by scaling the time direction in the symmetry algebra with a parameter $c$ and then taking a singular limit. The limit $c \rightarrow \infty$ describes the Galilean theory, whereas the limit $c \rightarrow 0$ gives the Carrollian theory. In this way, the symmetry algebra of non-relativistic theories is obtained from a relativistic parent theory by contraction. In supersymmetric theories, this procedure is also realized for supersymmetric transformation rules.

The Lagrangian is expected to provide a field-theoretic realization of the underlying symmetry algebra. In supersymmetric theories, however, the action of supersymmetry on the fields also introduces a multiplet structure that must be consistently realized in the Lagrangian. From this perspective, a non-relativistic contraction is a consistent group-theoretic operation that maps the algebraic relations into a distinct sector without truncating any charges. The same contraction can be implemented directly on the multiplet structure without spoiling its closure, since the contracted algebra remains consistent. It is then reflected in the Lagrangian as a decomposition into distinct sectors associated with the contracted multiplet or with its subsectors.

In this work, we consider off-shell supersymmetric Carrollian and Galilean scalar and vector field theories and show that a non-relativistic scaling decomposes the Lagrangian into distinct, individually consistent sectors, each realized by either the contracted multiplet or a truncated submultiplet. We present explicit three-dimensional examples, although the same mechanism applies more generally in other dimensions. The corresponding supersymmetric structures are obtained after the non-relativistic scaling
\begin{equation}
\mathcal{L}_{\mathrm{Rel}}
=
c^{n}\,\mathcal{L}_{\mathrm{Theory\,A}}
+\mathcal{L}_{\mathrm{Theory\,B}} \, .
\end{equation}
Here, $n$ denotes the scaling exponent, while $\mathcal{L}_{\text {Theory A }}$ and $\mathcal{L}_{\text {Theory B }}$ denote two sectors that may arise from the non-relativistic scaling. The sector $\mathcal{L}_{\text {Theory A }}$ is invariant under the contracted multiplet, whereas $\mathcal{L}_{\text {Theory B }}$ is invariant only under a contracted submultiplet. Thus, the non-relativistic limit can decompose the relativistic Lagrangian into distinct pieces, each of which is invariant under either the contracted multiplet or one of its subsectors, although more than two sectors may arise in a given scaling. 

\textit{$\mathcal{N}=2$ 2+1 Dimensional Twisted Carroll Theory}--
The three-dimensional twisted $\mathcal{N}=2$ Carroll superalgebra is generated by $J, P_a, C_a, H$, and $Q_\alpha^{ \pm}$, where $\alpha=1,2$ is a spinor index and $a=1,2$ is a spatial index, with the following non-vanishing (anti)commutators (see \cite{bulunur2026twistedoriginmagneticcarroll, Zorba_2025} for details).
\begin{equation} \label{SuperCarrollAlgebra}
\begin{aligned}
[J, P_a] &= \varepsilon_{ab} P^b, 
& [J, C_a] &= \varepsilon_{ab} C^b, \\[4pt]
[C_a, P_b] &= -\,\varepsilon_{ab} H, 
& [C_a, Q_\alpha^{+}] &= \tfrac12 (\gamma_a Q^{-})_\alpha, \\[4pt]
\{ Q_\alpha^{+}, Q_\beta^{+} \} 
  &= (\gamma^a C^{-1})_{\alpha\beta} P_a, 
& \{ Q_\alpha^{+}, Q_\beta^{-} \} 
  &= (\gamma^0 C^{-1})_{\alpha\beta} H, \\[4pt]
[J, Q_\alpha^{\pm}] 
  &= \tfrac12 (\gamma_0 Q^{\pm})_\alpha.
\end{aligned}
\end{equation}

\textit{Example 1. Scalar Theory}-- We begin with the relativistic off-shell real scalar theory
\begin{equation}
\begin{aligned}
\mathcal{L}_{Rel}= & \left(\partial_a \phi_1\right)^2-\left(\partial_a \phi_2\right)^2-2 F_1 \partial_0 \phi_2-2 F_2 \partial_0 \phi_1-F_1^2+F_2^2 \\
& +\bar{\chi}_{+} \gamma^a \partial_a \chi_{+}+\bar{\chi}_-\gamma^a \partial_a \chi_{-}+\bar{\chi}_+\gamma^0 \partial_0 \chi_{-}+\bar{\chi}_-\gamma^0 \partial_0 \chi_{+},
\end{aligned}
\end{equation}
where the theory consists of two real scalars $\phi_i$, two Majorana fermions $\chi_\pm$, and two real auxiliary fields $F_i$. This multiplet provides a convenient relativistic parent theory, and its supersymmetry transformations can be written in a $S O(1,1)$-covariant form as follows
\begin{equation}
\begin{aligned}
\delta \phi_1 &= \bar{\epsilon}_{+}\chi_{-}+\bar{\epsilon}_{-}\chi_{+},\\
\delta \phi_2 &= \bar{\epsilon}_{+}\gamma_0\chi_{+}-\bar{\epsilon}_{-}\gamma_0\chi_{-},\\
\delta \chi_{+} &=
-\gamma_0\bigl(\gamma^a\partial_a\phi_2+F_2\bigr)\epsilon_{+}
+\bigl(\gamma^a\partial_a\phi_1-2\partial_0\phi_2\\& \quad -F_1 \bigr)\epsilon_{-},\\
\delta \chi_{-} &=
\bigl(\gamma^a\partial_a\phi_1-F_1\bigr)\epsilon_{+}
+\bigl(2\gamma^0\partial_0\phi_1+\gamma_0\gamma^a\partial_a\phi_2\\& \quad+\gamma_0F_2\bigr)\epsilon_{-},\\
\delta F_1 &=
-\bar{\epsilon}_{+}\gamma^a\partial_a\chi_{-}
-\bar{\epsilon}_{-}\gamma^a\partial_a\chi_{+}
-2\bar{\epsilon}_{-}\gamma^0\partial_0\chi_{-},\\
\delta F_2 &=
-\bar{\epsilon}_{+}\gamma_0\gamma^a\partial_a\chi_{+}
+2\bar{\epsilon}_{-}\partial_0\chi_{+}
+\bar{\epsilon}_{-}\gamma_0\gamma^a\partial_a\chi_{-}.
\end{aligned}
\end{equation}
The model admits the following scalings
\begin{equation}
    \begin{aligned}
& \phi_1 \rightarrow c \phi_1, \quad \chi_{-} \rightarrow c \chi_{-}, \quad F_1 \rightarrow c F_1, \\
& \epsilon_{-} \rightarrow c \epsilon_{-},  \quad \partial_0 \rightarrow \frac{1}{c} \partial_0, \quad c \rightarrow0.
\end{aligned}
\end{equation}
Under the Carrollian scaling, the relativistic Lagrangian takes the form
\begin{equation}
    \mathcal{L}_{Rel}=c^2 \mathcal{L}_{(2)}+\mathcal{L}_{(0)},
\end{equation}
where we have 
\begin{equation}
    \begin{aligned}
        & \mathcal{L}_{(2)}=\left(\partial_a \phi_1\right)^2-F_1^2+\bar{\chi}_-\gamma^a \partial_a \chi_{-}, \\
        & \mathcal{L}_{(0)}=-\left(\partial_a \phi_2\right)^2-2 F_1 \partial_0 \phi_2-2 F_2 \partial_0 \phi_1+F_2^2 \\ & \qquad+\bar{\chi}_{+} \gamma^a \partial_a \chi_{+}+\bar{\chi}_{+} \gamma^0 \partial_0 \chi_{-}+\bar{\chi}_{-} \gamma^0 \partial_0 \chi_{+} .
    \end{aligned}
\end{equation}
and where $\mathcal{L}_{(0)}$ is invariant under the contracted multiplet
\begin{equation} \label{multiplet1}
    \begin{aligned}
&\delta \phi_1=\bar{\epsilon}_{+} \chi_{-}+\bar{\epsilon}_{-} \chi_{+} ,\\
&\delta \phi_2= \bar{\epsilon}_{+} \gamma_0 \chi_{+}, \\
&\delta \chi_{+}=\left(- \gamma_0 \gamma^a \partial_a \phi_2-\gamma_0 F_2\right) \epsilon_{+}-2 \left(\partial_0 \phi_2\right) \epsilon_{-}, \\
&\delta \chi_{-}=\left(\gamma^a \partial_a \phi_1- F_1\right) \epsilon_{+}+\left(2 \gamma^0 \partial_0 \phi_1+ \gamma_0 \gamma^a \partial_a \phi_2 +\gamma_0 F_2\right) \epsilon_{-}, \\
&\delta F_1=- \bar{\epsilon}_{+} \gamma^a \partial_a \chi_{-}- \bar{\epsilon}_{-} \gamma^a \partial_a \chi_{+}-2  \bar{\epsilon}_{-} \gamma^0 \partial_0 \chi_{-}, \\
&\delta F_2=-\bar{\epsilon}_{+} \gamma_0 \gamma^a \partial_a \chi_{+}+2 \bar{\epsilon}_{-} \partial_0 \chi_{+},
\end{aligned}
\end{equation}
whereas $\mathcal{L}_{(2)}$ is invariant under the submultiplet of eq. \eqref{multiplet1}, namely
\begin{equation}
    \begin{aligned}
&\delta \phi_1=\bar{\epsilon}_{+} \chi_{-}, \\
&\delta \chi_{-}=\left(\gamma^a \partial_a \phi_1-F_1\right) \epsilon_{+}, \\
&\delta F_1=-\bar{\epsilon}_{+} \gamma^a \partial_a \chi_{-} .
\end{aligned}
\end{equation}

\textit{Example 2. Vector Theory}--  The three-dimensional twisted $\mathcal{N}=2$ vector multiplet consists of a gauge field, a scalar, two spinors, and an auxiliary scalar, $(C_\mu,\rho,\lambda_\pm,D)$, with $\mu=0,1,2$. Using the three-dimensional Levi--Civita tensor with $\epsilon_{012}=+1$, we define the dual vector $V_\mu=\varepsilon_\mu{}^{\nu\rho}\partial_\nu C_\rho$, which satisfies $\partial^\mu V_\mu=0$. In terms of $(\rho,\lambda_\pm,V_\mu,D)$, the theory is described by
\begin{equation}
\begin{aligned}
\mathcal{L}_{\mathrm{Rel}}
&=\frac{1}{2} V_a^2-\frac{1}{2} V_0^2-\frac{1}{8}\left(\partial_0 \rho\right)^2+\frac{1}{8}\left(\partial_a \rho\right)^2-\frac{1}{2} D^2 \\
&\quad -\bar{\lambda}_{+} \gamma^a \partial_a \lambda_{+}
-\bar{\lambda}_{-} \gamma^a \partial_a \lambda_{-}
+\bar{\lambda}_{+} \gamma_0 \partial_0 \lambda_{-}
+\bar{\lambda}_{-} \gamma_0 \partial_0 \lambda_{+},
\end{aligned}
\end{equation}
with the following supersymmetry transformation rules
\begin{equation}
\begin{aligned}
\delta \rho
&=-2\Big(\bar{\epsilon}_{+}\gamma_{0}\lambda_{+}-\bar{\epsilon}_{-}\gamma_{0}\lambda_{-}\Big),
\\[4pt]
\delta \lambda_{+}
&=-\frac{1}{2}\gamma^{a}\epsilon_{-}V_{a}
+\frac{1}{2}\gamma_{0}\epsilon_{+}V_{0}
+\frac{1}{2}\gamma_{0}\epsilon_{-}D
\\ &+\frac{1}{4}\gamma_{0}\gamma^{a}\epsilon_{+}\partial_{a}\rho
+\frac{1}{4}\epsilon_{-}\partial_{0}\rho,
\\[4pt]
\delta \lambda_{-}
&=-\frac{1}{2}\gamma^{a}\epsilon_{+}V_{a}
+\frac{1}{2}\gamma_{0}\epsilon_{-}V_{0}
-\frac{1}{2}\gamma_{0}\epsilon_{+}D
\\ &-\frac{1}{4}\gamma_{0}\gamma^{a}\epsilon_{-}\partial_{a}\rho
-\frac{1}{4}\epsilon_{+}\partial_{0}\rho,
\\[4pt]
\delta D
&=-\bar\epsilon_{+}\,\partial_{0}\lambda_{+}
-\bar\epsilon_{+}\,\gamma_{0}\gamma^{a}\,\partial_{a}\lambda_{-}
+\bar\epsilon_{-}\,\partial_{0}\lambda_{-}
\\ & +\bar\epsilon_{-}\,\gamma_{0}\gamma^{a}\,\partial_{a}\lambda_{+},
\\[4pt]
\delta V_{0}
&=\bar{\epsilon}_{+}\gamma_{0}{}^{a}\partial_{a}\lambda_{+}
+\bar{\epsilon}_{-}\gamma_{0}{}^{a}\partial_{a}\lambda_{-},
\\[4pt]
\delta V_{a}
&=\bar{\epsilon}_{+}\gamma_{a}{}^{0}\partial_{0}\lambda_{+}
+\bar{\epsilon}_{-}\gamma_{a}{}^{0}\partial_{0}\lambda_{-}
+\bar{\epsilon}_{-}\gamma_{a}{}^{b}\partial_{b}\lambda_{+} \\ &
+\bar{\epsilon}_{+}\gamma_{a}{}^{b}\partial_{b}\lambda_{-},
\end{aligned}
\end{equation}
we consider the scaling
\begin{equation}
\begin{aligned}
     & \rho \rightarrow c  \rho, \quad \lambda_+ \rightarrow c  \lambda_+, \quad   V_0 \rightarrow c  V_0,\\
& \epsilon_- \rightarrow c \epsilon_- ,\quad   \partial_0 \rightarrow \frac{1}{c} \partial_0, \quad c \rightarrow0,
\end{aligned}
\end{equation}
then the Lagrangian decomposes under this scaling as
\begin{equation}
    \mathcal{L}_{\text {Rel }}=c^2 \mathcal{L}_{(2)}+\mathcal{L}_{(0)},
\end{equation}
where we have
\begin{equation}
\begin{aligned}
       & \mathcal{L}_{\mathrm{(2)}}=-\frac{1}{2} V_0^2+\frac{1}{8}\left(\partial_a \rho\right)^2-\bar{\lambda}_{+} \gamma^a \partial_a \lambda_{+} ,\\
        & \mathcal{L}_{\mathrm{(0)}}=\frac{1}{2} V_a ^2-\frac{1}{8}\left(\partial_0 \rho\right)^2-\frac{1}{2} D^2 \\ &-\bar{\lambda}_{-} \gamma^a \partial_a \lambda_{-}+\bar{\lambda}_{+} \gamma_0 \partial_0 \lambda_{-}+\bar{\lambda}_{-} \gamma_0 \partial_0 \lambda_{+}.
\end{aligned}
\end{equation} 
$\mathcal{L}_{(0)}$ is invariant under the contracted multiplet
\begin{equation}
\begin{aligned}
\delta \rho
&=-2\Big(\bar{\epsilon}_{+}\gamma_{0}\lambda_{+}-\bar{\epsilon}_{-}\gamma_{0}\lambda_{-}\Big),
\\[4pt]
\delta \lambda_{+}
&=-\frac{1}{2}\gamma^{a}\epsilon_{-}V_{a}
+\frac{1}{2}\gamma_{0}\epsilon_{+}V_{0}
+\frac{1}{2}\gamma_{0}\epsilon_{-}D \\ & +\frac{1}{4}\gamma_{0}\gamma^{a}\epsilon_{+}\partial_{a}\rho
+\frac{1}{4}\epsilon_{-}\partial_{0}\rho,
\\[4pt]
\delta \lambda_{-}
&=-\frac{1}{2}\gamma^{a}\epsilon_{+}V_{a}
-\frac{1}{2}\gamma_{0}\epsilon_{+}D
-\frac{1}{4}\epsilon_{+}\partial_{0}\rho,
\\[4pt]
\delta D
&=-\bar\epsilon_{+}\,\partial_{0}\lambda_{+}
-\bar\epsilon_{+}\,\gamma_{0}\gamma^{a}\,\partial_{a}\lambda_{-}
+\bar\epsilon_{-}\,\partial_{0}\lambda_{-},
\\[4pt]
\delta V_{0}
&=\bar{\epsilon}_{+}\gamma_{0}{}^{a}\partial_{a}\lambda_{+}
+\bar{\epsilon}_{-}\gamma_{0}{}^{a}\partial_{a}\lambda_{-},
\\[4pt]
\delta V_{a}
&=\bar{\epsilon}_{+}\gamma_{a}{}^{0}\partial_{0}\lambda_{+}
+\bar{\epsilon}_{-}\gamma_{a}{}^{0}\partial_{0}\lambda_{-}
+\bar{\epsilon}_{+}\gamma_{a}{}^{b}\partial_{b}\lambda_{-},
\end{aligned}
\end{equation}
while $\mathcal{L}_{(2)}$ is invariant under the following submultiplet
\begin{equation}
    \begin{aligned}
\delta \rho & =-2 \bar{\epsilon}_{+} \gamma_0 \lambda_{+} \\
\delta \lambda_{+} & =\frac{1}{2} \gamma_0 \epsilon_{+} V_0+\frac{1}{4} \gamma_0 \gamma^a \epsilon_{+} \partial_a \rho, \\
\delta V_0 & =\bar{\epsilon}_{+} \gamma_0{ }^a \partial_a \lambda_{+}.
\end{aligned}
\end{equation}
We now turn to the Galilean case, which arises from the $c\to\infty$ contraction.

\textit{$\mathcal{N}=2$ 2+1 Dimensional Galilean Theory}--
The three-dimensional $\mathcal{N}=2$ Galilean superalgebra is generated by $J$, $P_a$, $G_a$, $H$, and $Q_\alpha^\pm$, with the following non-vanishing (anti)commutators \cite{Bergshoeff:2015uaa}
\begin{equation}
\begin{aligned} {\left[J, P_a\right] } & =\varepsilon_{a b} P_b, \quad\left[J, G_a\right]=\varepsilon_{a b} G_b, \\ {\left[G_a, H\right] } & =-\varepsilon_{a b} P_b, \quad\left[G_a, Q^{+}\right]=-\gamma_a Q^{-}, \\ {\left[J, Q^{ \pm}\right] } & =-\frac{1}{2} \gamma_0 Q^{ \pm}, \quad \{Q_\alpha^{+}, Q_\beta^{+}\}  =2\left(\gamma^0 C^{-1}\right)_{\alpha \beta} H, \\ \{Q_\alpha^{+}, Q_\beta^{-}\} & =(\gamma^a C^{-1})_{\alpha \beta} P_a.
\end{aligned}
\end{equation}

\textit{Example 3. Scalar Theory}-- The theory consists of two real scalars $\phi_i$, two Majorana fermions $\chi_\pm $, and two real auxiliary fields $F_i$. 
\begin{equation}
\begin{aligned}
\mathcal{L}_{Rel}
&= -\left(\partial_0 \phi_1\right)^2-\left(\partial_0 \phi_2\right)^2
+\left(\partial_a \phi_1\right)^2+\left(\partial_a \phi_2\right)^2-\frac{1}{4} F_1^2 , \\
&\quad  -\frac{1}{4} F_2^2 -\bar{\chi}_+\gamma_0 \partial_0 \chi_{+}
-\bar{\chi}_{-} \gamma_0 \partial_0 \chi_{-}
+\bar{\chi}_{+} \gamma^a \partial_a \chi_{-}
 \\ &\qquad +\bar{\chi}_{-} \gamma^a \partial_a \chi_{+},
\end{aligned}
\end{equation}
with the following multiplet structure
\begin{equation}
\begin{aligned}
& \delta \phi_1=\bar{\epsilon}_{+} \chi_{+}+\bar{\epsilon}_{-} \chi_{-}, \\
& \delta \phi_2=-\bar{\epsilon}_{+} \gamma_0 \chi_{+}+\bar{\epsilon}_{-} \gamma_0 \chi_{-}, \\
& \delta \chi_{+}=\left(-\gamma_0 \partial_0 \phi_1+\partial_0 \phi_2\right) \epsilon_{+} \\
& \qquad +\left(\gamma^a \partial_a \phi_1+\gamma_0 \gamma^a \partial_a \phi_2-\frac{1}{2} F_1-\frac{1}{2} \gamma_0 F_2\right) \epsilon_{-}, \\
& \delta \chi_{-}=\left(\gamma^a \partial_a \phi_1-\gamma_0 \gamma^a \partial_a \phi_2-\frac{1}{2} F_1+\frac{1}{2} \gamma_0 F_2\right) \epsilon_{+} \\
&\qquad  +\left(-\gamma_0 \partial_0 \phi_1-\partial_0 \phi_2\right) \epsilon_{-} \text {, } \\
& \delta F_1=2 \bar{\epsilon}_{+}\left(\gamma_0 \partial_0-\gamma^a \partial_a\right) \chi_{-}+2 \bar{\epsilon}_{-}\left(\gamma_0 \partial_0-\gamma^a \partial_a\right) \chi_{+}, \\
& \delta F_2=-2 \bar{\epsilon}_{+}\left(\partial_0 \chi_{-}+\gamma_0 \gamma^a \partial_a \chi_{+}\right)+2 \bar{\epsilon}_{-}\left(\partial_0 \chi_{+}+\gamma_0 \gamma^a \partial_a \chi_{-}\right),
\end{aligned}
\end{equation}
we impose the scaling
\begin{equation}
\begin{aligned}
  &  \epsilon_{+} \to c^{1/2}\,\epsilon_{+}, \ \
\epsilon_{-} \to c^{-1/2}\,\epsilon_{-}, \ \
\partial_{0} \to \frac{1}{c}\,\partial_{0}, \ \
\chi_+ \rightarrow c^{\frac{1}{2}} \chi_+ ,\\ 
&\chi_- \rightarrow c^{\frac{3}{2}} \chi_-, \quad \phi_i \rightarrow c \phi_i, \quad F_i \rightarrow c F_i, \quad c \to \infty.
\end{aligned}
\end{equation}
The Lagrangian takes the following form
\begin{equation}
\mathcal{L}_{Rel}=\mathcal{L}_{(0)}+c^2 \mathcal{L}_{(2)},
\end{equation}
where we have
\begin{equation}
    \begin{aligned}
&\mathcal{L}_{(0)}
= -\left(\partial_0 \phi_1\right)^2-\left(\partial_0 \phi_2\right)^2-\bar{\chi}_+\gamma_0 \partial_0 \chi_{+}, \\
&\mathcal{L}_{(2)}=
\left(\partial_a \phi_1\right)^2+\left(\partial_a \phi_2\right)^2-\bar{\chi}_{-} \gamma_0 \partial_0 \chi_{-}+\bar{\chi}_{+} \gamma^a \partial_a \chi_{-} \\ 
&+\bar{\chi}_{-} \gamma^a \partial_a \chi_{+}-\frac{1}{4} F_1^2-\frac{1}{4} F_2^2.
    \end{aligned}
\end{equation}
$\mathcal{L}_{(0)}$ is invariant under the contracted multiplet
\begin{equation}
\begin{aligned}
\delta \phi_1= & \bar{\epsilon}_{+} \chi_{+}+\bar{\epsilon}_{-} \chi_{-}, \\
\delta \phi_2= & -\bar{\epsilon}_{+} \gamma_0 \chi_{+}+\bar{\epsilon}_{-} \gamma_0 \chi_{-}, \\
\delta \chi_{+}= & \left(-\gamma_0 \partial_0 \phi_1+\partial_0 \phi_2\right) \epsilon_{+} \\
& \quad+\left(\gamma^a \partial_a \phi_1+\gamma_0 \gamma^a \partial_a \phi_2-\frac{1}{2} F_1-\frac{1}{2} \gamma_0 F_2\right) \epsilon_{-}, \\
\delta \chi_{-}= & \left(\gamma^a \partial_a \phi_1-\gamma_0 \gamma^a \partial_a \phi_2-\frac{1}{2} F_1+\frac{1}{2} \gamma_0 F_2\right) \epsilon_{+}, \\
\delta F_1= & 2 \bar{\epsilon}_{+}\left(\gamma_0 \partial_0 \chi_{-}-\gamma^a \partial_a \chi_{+}\right)-2 \bar{\epsilon}_{-} \gamma^a \partial_a \chi_{-}, \\
\delta F_2= & -2 \bar{\epsilon}_{+}\left(\partial_0 \chi_{-}+\gamma_0 \gamma^a \partial_a \chi_{+}\right)+2 \bar{\epsilon}_{-} \gamma_0 \gamma^a \partial_a \chi_{-} ,
\end{aligned}
\end{equation}
whereas $\mathcal{L}_{(2)}$ is invariant under the submultiplet 
\begin{equation}
\begin{aligned}
\delta \phi_1&=\bar{\epsilon}_{+}\chi_{+},\\[4pt]
\delta \phi_2&=-\bar{\epsilon}_{+}\gamma_0\chi_{+},\\[6pt]
\delta \chi_{+}&=\left(-\gamma_0\partial_0\phi_1+\partial_0\phi_2\right)\epsilon_{+},\\[6pt]
\delta \chi_{-}&=\left(\gamma^a\partial_a\phi_1-\gamma_0\gamma^a\partial_a\phi_2-\frac12 F_1+\frac12\gamma_0F_2\right)\epsilon_{+},\\[6pt]
\delta F_1&=2\bar{\epsilon}_{+}\left(\gamma_0\partial_0\chi_{-}-\gamma^a\partial_a\chi_{+}\right),\\[6pt]
\delta F_2&=-2\bar{\epsilon}_{+}\left(\partial_0\chi_{-}+\gamma_0\gamma^a\partial_a\chi_{+}\right).
\end{aligned}
\end{equation}
\\ 

\textit{Example 4. Vector Theory}-- The three-dimensional $\mathcal{N}=2$ vector multiplet is described by a scalar $\rho$, two spinors $\lambda_\pm$, the spatial vector components $V_a$, and the temporal and auxiliary fields $V_0$ and $D$. For the non-relativistic contraction, it is convenient to replace $V_0$ and $D$ by the linear combinations
\begin{equation}
    U=V_0+D,\qquad W=V_0-D.
\end{equation}
In terms of the variables $(\rho,\lambda_\pm,V_a,U,W)$, the relativistic Lagrangian is given by
\begin{equation}
\begin{aligned}
\mathcal{L}_{\mathrm{Rel}}&=  \frac{1}{2} V_a^2-\frac{1}{2} U W+\frac{1}{8}\left(\partial_0 \rho\right)^2-\frac{1}{8}\left(\partial_a \rho\right)^2+\bar{\lambda}_{+} \gamma_0 \partial_0 \lambda_{+} \\
& +\bar{\lambda}_{-} \gamma_0 \partial_0 \lambda_{-}-\bar{\lambda}_{+} \gamma^a \partial_a \lambda_{-}-\bar{\lambda}_{-} \gamma^a \partial_a \lambda_{+} .
\end{aligned}
\end{equation}
with
\begin{equation}
\begin{aligned}
\delta \rho
&=-2\,\bar{\epsilon}_{+}\gamma_{0}\lambda_{+}
+2\,\bar{\epsilon}_{-}\gamma_{0}\lambda_{-},
\\[4pt]
\delta \lambda_{+}
&=-\frac{1}{2}\gamma^{a}\epsilon_{-}V_{a}
+\frac{1}{2}\gamma_{0}\epsilon_{+}U
+\frac{1}{4}\epsilon_{+}\partial_{0}\rho
+\frac{1}{4}\gamma_{0}\gamma^{a}\epsilon_{-}\partial_{a}\rho,
\\[4pt]
\delta \lambda_{-}
&=-\frac{1}{2}\gamma^{a}\epsilon_{+}V_{a}
+\frac{1}{2}\gamma_{0}\epsilon_{-}W
-\frac{1}{4}\epsilon_{-}\partial_{0}\rho
-\frac{1}{4}\gamma_{0}\gamma^{a}\epsilon_{+}\partial_{a}\rho,
\\[4pt]
\delta U
&=-\bar{\epsilon}_{+}\partial_{0}\lambda_{+}
+\bar{\epsilon}_{-}\partial_{0}\lambda_{-}
+2\,\bar{\epsilon}_{-}\gamma_{0}\gamma^{a}\partial_{a}\lambda_{+},
\\[4pt]
\delta W
&=\bar{\epsilon}_{+}\partial_{0}\lambda_{+}
+2\,\bar{\epsilon}_{+}\gamma_{0}\gamma^{a}\partial_{a}\lambda_{-}
-\bar{\epsilon}_{-}\partial_{0}\lambda_{-},
\\[4pt]
\delta V_{a}
&=-\bar{\epsilon}_{+}\gamma_{a}\gamma_{0}\partial_{0}\lambda_{-}
+\bar{\epsilon}_{+}\gamma_{a}{}^{b}\partial_{b}\lambda_{+}
-\bar{\epsilon}_{-}\gamma_{a}\gamma_{0}\partial_{0}\lambda_{+}
+\bar{\epsilon}_{-}\gamma_{a}{}^{b}\partial_{b}\lambda_{-}.
\end{aligned}
\end{equation}
The theory admits the following contraction
\begin{equation}
\begin{aligned}
&\epsilon_{+} \rightarrow c^{1 / 2} \epsilon_{+}, \ \ \epsilon_{-} \rightarrow c^{-1 / 2} \epsilon_{-}, \ \ \partial_0 \rightarrow \frac{1}{c} \partial_0, \ \  \\
& \lambda_{+} \rightarrow c^{1 / 2} \lambda_{+}, \quad \lambda_{-} \rightarrow c^{3 / 2} \lambda_{-}, \quad \rho \rightarrow c \rho,  \\ 
&V_a \rightarrow c V_a, \quad W \rightarrow c^2 W, \quad c \to \infty.
\end{aligned}
\end{equation}
The Lagrangian is scaled as
\begin{equation}
    \mathcal{L}_{\text {Rel }}=\mathcal{L}_{(0)}+c^2 \mathcal{L}_{(2)},
\end{equation}
where we have
\begin{equation}
\begin{aligned}
&\mathcal{L}_{(2)}=\frac{1}{2} V_a ^2-\frac{1}{2} U W-\frac{1}{8}\left(\partial_a \rho\right)^2 \\&+\bar{\lambda}_{-} \gamma_0 \partial_0 \lambda_{-} -\bar{\lambda}_{+} \gamma^a \partial_a \lambda_{-}-\bar{\lambda}_{-} \gamma^a \partial_a \lambda_{+} ,\\
&\mathcal{L}_{(0)}=\frac{1}{8}\left(\partial_0 \rho\right)^2+\bar{\lambda}_{+} \gamma_0 \partial_0 \lambda_{+},
\end{aligned}
\end{equation}
and $ \mathcal{L}_{(2)}$ is invariant under the contracted multiplet
\begin{equation}
\begin{aligned}
\delta \rho
&=-2\,\bar{\epsilon}_{+}\gamma_{0}\lambda_{+}
+2\,\bar{\epsilon}_{-}\gamma_{0}\lambda_{-},
\\
\delta \lambda_{+}
&=-\frac{1}{2}\gamma^{a}\epsilon_{-}V_{a}
+\frac{1}{2}\gamma_{0}\epsilon_{+}U
+\frac{1}{4}\epsilon_{+}\partial_{0}\rho
+\frac{1}{4}\gamma_{0}\gamma^{a}\epsilon_{-}\partial_{a}\rho,
\\
\delta \lambda_{-}
&=-\frac{1}{2}\gamma^{a}\epsilon_{+}V_{a}
+\frac{1}{2}\gamma_{0}\epsilon_{-}W
-\frac{1}{4}\gamma_{0}\gamma^{a}\epsilon_{+}\partial_{a}\rho,
\\
\delta U
&=-\bar{\epsilon}_{+}\partial_{0}\lambda_{+}
+\bar{\epsilon}_{-}\partial_{0}\lambda_{-}
+2\,\bar{\epsilon}_{-}\gamma_{0}\gamma^{a}\partial_{a}\lambda_{+},
\\
\delta W
&=2\,\bar{\epsilon}_{+}\gamma_{0}\gamma^{a}\partial_{a}\lambda_{-},
\\
\delta V_{a}
&=-\bar{\epsilon}_{+}\gamma_{a}\gamma_{0}\partial_{0}\lambda_{-}
+\bar{\epsilon}_{+}\gamma_{a}{}^{b}\partial_{b}\lambda_{+}
+\bar{\epsilon}_{-}\gamma_{a}{}^{b}\partial_{b}\lambda_{-},
\end{aligned}
\end{equation}
while $\mathcal{L}_{(0)}$ is invariant under
\begin{equation}
\begin{aligned}
&\delta \rho  =-2 \bar{\epsilon}_{+} \gamma_0 \lambda_{+} ,\\
&\delta \lambda_{+}  =\left(\frac{1}{2} \gamma_0 U+\frac{1}{4} \partial_0 \rho\right) \epsilon_{+}, \\
&\delta U  =-\bar{\epsilon}_{+} \partial_0 \lambda_{+}.
\end{aligned}
\end{equation}

It is important to note that, in the last case, the condition $\partial_0 U=0$ follows from the dual-vector constraint $\partial^\mu V_\mu=0$. After the contraction and subsequent truncation to the submultiplet, this relation reduces precisely to $\partial_0 U=0$ and is required for the off-shell invariance of the Lagrangian. 

\textit{Conclusion}-- In this work, we have shown that the non-relativistic contraction decomposes the theory into distinct sectors, each of which is invariant under either the contracted multiplet or one of its submultiplets. This perspective may be useful for the construction and analysis of electric and magnetic non-relativistic theories.

\textit{Acknowledgements}--
I am grateful to the theory group at Istanbul Technical University, especially Mehmet Ozkan
, Ilayda Bulunur, Mustafa Salih Zog, and Oguzhan Kasikci, for valuable discussions.

	\bibliographystyle{utphys}
	\bibliography{ref}

\end{document}